\documentclass[10pt]{article}
\usepackage{geometry}
\usepackage[utf8]{inputenc}

\usepackage[backend=biber,style=chem-acs,maxnames=99,giveninits=true, sorting=none]{biblatex}
\DeclareDelimFormat[bib,biblist]{multicitedelim}{\addcomma} 
\DeclareDelimFormat{multicitedelim}{\addcomma} 
\renewcommand{\cite}[1]{\textsuperscript{\autocite{#1}}}

\DeclareBibliographyDriver{article}{%
  \usebibmacro{author}%
  \setunit{\labelnamepunct}\newblock
  \usebibmacro{title}%
  \newblock
  \printfield{journaltitle}%
  \setunit{\space}%
  \textbf{\printfield{year}}%
  \setunit*{\addcomma\space}%
  \printfield{volume}%
  \setunit*{\addcomma\space}%
  \printfield{pages}%
  \setunit{\finentry\\}%
  \printfield{doi}%
  \finentry
}
\DeclareBibliographyDriver{misc}{%
  \usebibmacro{author} 
  \setunit{\labelnamepunct}\newblock
  \usebibmacro{title}%
  \newblock
  \textit{, Preprint} 
  \setunit{\addspace}%
  \textbf{\printfield{year}} 
  \setunit{\finentry\\}%
  \printfield{url}%
  \finentry
}
\DeclareBibliographyDriver{inproceedings}{%
  \usebibmacro{author}%
  \setunit{\labelnamepunct}\newblock
  \textnormal{\usebibmacro{title}}%
  \newblock
  \printfield{booktitle}%
  \setunit{\space}%
  \textbf{\printfield{year}}%
  \setunit*{\addcomma\space}%
  \printfield{volume}%
  \setunit*{\addcomma\space}%
  \printfield{pages}%
  \finentry
}

\usepackage{hyperref}
\hypersetup{
    colorlinks,
    citecolor=black,
    filecolor=black,
    linkcolor=black,
    urlcolor=black
}

\usepackage{titlesec}
\usepackage{stmaryrd}
\usepackage{float}
\usepackage{soul}
\usepackage{amsmath}

\usepackage{subcaption}
\usepackage[export]{adjustbox}
\usepackage{wrapfig}
\usepackage{graphicx}
\graphicspath{ {./initstruc_vs_ref/} }
    
\usepackage{booktabs}
\usepackage{dcolumn}
\newcolumntype{d}[1]{D..{#1}}
\usepackage{array}
\usepackage{stmaryrd}
\SetSymbolFont{stmry}{bold}{U}{stmry}{b}{n}

\title{Short-range $\Delta$-Machine Learning: A cost-efficient strategy to transfer chemical accuracy to condensed phase systems}

\graphicspath{{./figures_SI}}
\graphicspath{{./figures}}
\usepackage{multicol}
\usepackage{balance}
\usepackage{titlesec}
\titlespacing{\section}{0pt}{10pt}{5pt}
\usepackage{setspace}
\begin{document}
\newcommand{\mbscan}[0]{\mbox{MB-SCAN}}
\newcommand{\mbpol}[0]{\mbox{MB-pol}}
\newcommand{\dmlp}[0]{sr$\Delta$ML@cluster}
\newcommand{\bmlp}[0]{ML@PBC}
\newcommand{\rmsef}[0]{RMSE($\boldsymbol{F}$)}
\newcommand{\mevpera}[0]{meV$\cdot$A$^{-1}$}
\newcommand{\mevperatom}[0]{meV$\cdot$atom$^{-1}$}
\newcommand{\deltac}[0]{$\Delta_\mathrm{SCAN}^\mathrm{CC}$}
\newcommand{\deltaace}[0]{ACE($\Delta_\mathrm{SCAN}^\mathrm{CC}$)}
\newcommand{\baselineace}[0]{ACE(SCAN)}
\newcommand{\clusterdelta}[0]{sr$\Delta$ML}
\newcommand{\nwater}[1]{(H$_2$O)$_{#1}$}

\date{}
\author{}
\maketitle
\vspace{-5em}
\begin{center}
\large
Bence Balázs Mészáros$^{1,2}$, András Szabó$^2$, János Daru$^{2,}$*\\
\vspace{1em}
\small\textit{
Hevesy György PhD School of Chemistry Eötvös Loránd University, Pázmány Péter sétány 1/A, 1117 Budapest, Hungary\\
Department of Organic Chemistry, Eötvös Loránd University, Pázmány Péter sétány 1/A, 1117 Budapest, Hungary\\}
\vspace{1em}
E-mail: janos.daru@ttk.elte.hu
\end{center}

\begin{multicols}{2}
\setlength{\parskip}{0pt}
\section{Abstract}

DFT-based machine-learning potentials (MLPs) are now routinely trained for condensed-phase systems, but surpassing DFT accuracy remains challenging due to the cost or unavailability of periodic reference calculations. Our previous work (PRL 2022, 129, 226001) demonstrated that high-accuracy periodic MLPs can be trained within the CCMD framework using extended yet finite reference calculations. Here, we introduce \textit{short-range $\Delta$-Machine Learning} (\clusterdelta), which builds on periodic MLPs while accurately reproducing the observables of the high-level method.
\vspace{2em}
\\
\begin{figure}[H]
    \centering
    \includegraphics[width=1.0\linewidth]{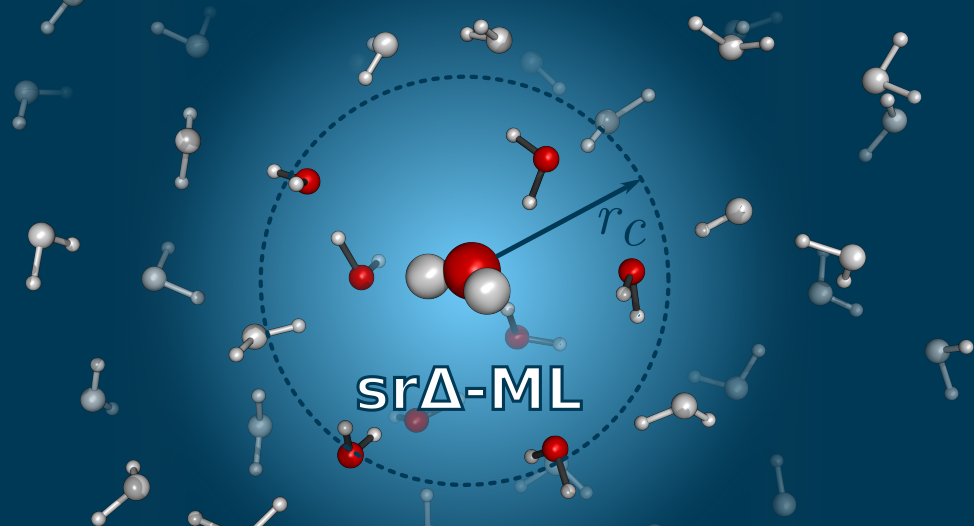}
\end{figure}
\end{multicols}
\hrule
\begin{multicols}{2}

\section{Introduction}
The advancement of machine-learning potentials (MLP) in the past decade enabled large-scale simulations of condensed-phase systems with high fidelity to the applied reference method. State-of-the-art frameworks enable the nanosecond-long simulation of molecules or periodic systems containing thousands to millions of atoms with excellent accuracy and data efficiency, many of them relying on the Behler-Parinello symmetry functions,\cite{Behler2007-fo, Behler2011-nn, Ko2021-ob, Daru2022-rj, Shanavas-Rasheeda2022-ko, O-Neill2024-wv, Kocer2025-zs} the Smooth Overlap of Atomic Positions (SOAP),\cite{Bartok2013-yp, Cliffe2017-zh, Dragoni2018-mx, Deringer2021-yk, Stocker2023-by, Schaaf2023-fs, Wan2025-ma} or the Atomic Cluster Expansion (ACE)\cite{Drautz2019-sg, Lysogorskiy2021-dx, Batatia2022-cf, Kovacs2023-ab, Kovacs2023-qa, Cheng2024-vb, Bochkarev2024-bv, Kaur2024-bf} to represent the local atomic structure. Many models employ message-passing neural networks (MPNN) to describe non-local interactions.\cite{Batatia2022-cf, Frank2024-xe, Unke2021-xf, Batzner2022-es, Musaelian2023-jf, Maxson2024-ij, Geiger2022-mz, Chen2024-yk, Schutt2021-zz} Applications include high-performance potentials with DFT accuracy for covalent materials\cite{Ibrahim2024-bu, Lysogorskiy2021-dx}, soft matter,\cite{Qamar2023-yn, Veit2019-kd} molten salts\cite{Finkbeiner2024-rv, Batzner2022-es}, metals\cite{Dragoni2018-mx,Lysogorskiy2021-dx,Bochkarev2022-wz, Ibrahim2023-vg}, general-purpose organic force-fields,\cite{Kovacs2023-ab, Smith2019-tt} crystal structure prediction of perovskites\cite{Karimitari2024-qt}, reactive condensed-phase simulations,\cite{Zhang2024-qt, Celerse2024-ut} and very recently, biomolecular dynamics.\cite{Unke2024-aw, Wang2024-lg}  However, obtaining high-quality and representative reference data for training remains a significant challenge.\cite{Poltavsky2025-kj} One common approach for generating training data is \textit{ab initio} molecular dynamics (AIMD),\cite{Marx2000-ex} which can treat complex systems, such as solutes and surfaces while providing dynamical properties. However, the high computational cost of these simulations necessitates compromises in sampling time and the level of theory applied. Consequently, AIMD-based MLPs typically use pure functionals (GGA or mGGA) as the ground truth.\cite{Dragoni2018-mx, Schran2020-cc, Yao2020-yv, Lysogorskiy2021-dx, Qamar2023-yn, Mondal2023-bw, Finkbeiner2024-rv, Ibrahim2024-bu, Karimitari2024-qt, Li2024-uw, Stolte2025-nl} This level can give a satisfactory description of some periodic materials but is known to include discrepancies due to their inadequate electronic-structure description.\cite{Marsalek2017-mw, Ruiz-Pestana2018-fd, Duignan2020-an, Zhang2021-xk, Dasgupta2022-zg, Li2024-uw, O-Neill2024-wv} To surpass these limitations, higher-level empirical or \textit{ab initio} PBC calculations were also used as a reference for the training of MLPs. This approach was demonstrated for hybrid,\cite{Zhang2021-xk} double hybrid, MP2,\cite{O-Neill2024-wv} RPA,\cite{Yao2021-bl, O-Neill2024-wv} and diffusion Monte Carlo\cite{Tirelli2022-yw, Slootman2024-vc, Goswami2025-ls} references. However, their prohibitive cost and the lack of systematic improvability hinder the broader spread of this methodology.

Meanwhile, systematically improvable quantum chemistry methods, such as the gold-standard of quantum chemistry CCSD(T), are limited to gas-phase\cite{Szabo2015-bb, Stei2016-xn, Chmiela2017-ph, Schran2018-py, Chmiela2018-ej, Sauceda2019-jj, Schran2020-wu, Gyori2020-ri, Fu2020-bj, Smith2020-ys, Meyer2021-wz, Zheng2021-dj, Kaser2022-fk, Kim2024-tm, Nandi2024-tb, Czako2024-mz} or require compromises in the size of the periodic images. For example, bulk water represented by 16 molecules,\cite{Chen2023-la}, or a zeolite containing 40 atoms were modeled by MLPs trained on PBC coupled cluster data.\cite{Herzog2024-st} On the contrary, for finite systems the local correlation variants of CCSD(T) such as the domain-based local natural orbital (DLPNO)\cite{Guo2018-kc, Liakos2020-bv} or the local natural orbital (LNO)\cite{Nagy2017-me, Nagy2018-mp, Nagy2019-lt, Al-Hamdani2021-tq} approximations of CCSD(T) offer promising scalability.

Building on these new advances, we recently proposed the coupled cluster molecular dynamics (CCMD) protocol,\cite{Daru2022-rj, Stolte2024-tj} where we trained an MLP on cluster data and applied it to periodic simulations (cluster-to-bulk strategy\cite{Wilmer2012-dj, Alireza-Ghasemi2015-zn, Gastegger2016-zz, Faraji2017-gi, Eckhoff2019-sl, Yu2022-qk, Zaverkin2022-fk, Zhang2024-qt}). To this end, we have applied a two-step delta-learning procedure. The first step involved learning the difference between an analytical baseline potential and DLPNO-MP2, while the second one accounted for the difference between local MP2 and local coupled cluster. The final coupled cluster level potential was used to run PBC water simulations, incorporating nuclear quantum effects (NQEs) through path-integral molecular dynamics (PIMD), giving excellent agreement with experimental data. Although this approach enables cluster-to-bulk learning at the gold-standard level of quantum chemistry, a large number of reference calculations (3,000 DLPNO-CCSD(T) and over 13,000 DLPNO-MP2 for \nwater{64} clusters) resulted in a substantial computational price. While the examples mentioned above demonstrated promising results, the training was often done on large clusters, and the lack of periodic reference data prohibited the direct validation of the resulting potentials' performance in the cluster-to-bulk transfer.

So far, we have seen two major approaches for training MLPs to simulate condensed phase systems. The first, routinely applied approach involves bulk data in the training set to predict the forces and energy of a periodic system, termed $E_\mathrm{ML@PBC}(R_\mathrm{PBC})$. The alternative approach is to use MLPs trained solely on cluster data to predict bulk energy $E_\mathrm{ML@cluster}(R_\mathrm{PBC})$. In this work, we propose the \clusterdelta\ procedure combining a baseline-MLP trained on periodic data (\bmlp) at a lower level and a short-ranged $\Delta$-potential trained on clusters for the difference between the higher and the lower level method (\dmlp). Accordingly, the final PBC energy expression is described in the following equation, 
\begin{multline}
        E_\mathrm{ML}(\boldsymbol{R}_\mathrm{PBC}) =\\
        E_{\mathrm{ML@PBC}}(\boldsymbol{R}_\mathrm{PBC}) + E_\mathrm{sr\Delta ML@cluster}(\boldsymbol{R}_\mathrm{PBC}),
\end{multline}

is obtained by the sum of two independently trained MLPs.

A special requirement for the lower-level method is the consistent implementations of energies and forces for both clusters and periodic systems, which is readily available for many state-of-the-art methods, such as tight-binding, DFT, HF, MP2, and RPA. Meanwhile,  the higher-level method for cluster calculations can be chosen with great flexibility, therefore enabling potentials based on cutting-edge methodologies without PBC implementation and the need for analytical forces.

Since the construction of \bmlp\ can be done routinely and data-efficiently using modern ML frameworks, this work focuses on training a \dmlp\ model. While this approach is not constrained to a specific baseline model, we aimed to choose one that provides a good cost-performance balance in AIMD simulations, the meta-GGA SCAN.\cite{Sun2015-ke, Sun2016-xh} Since the validation of the final potential is not feasible due to the lack of large-scale PBC CC simulations, we chose the highly accurate many-body expansion-based surrogate surfaces, \mbscan,\cite{Dasgupta2021-uh, Dasgupta2022-zg} and \mbpol\cite{Babin2013-tp, Medders2014-lw, Babin2014-fd, Reddy2016-pa, Paesani2016-wy, Zhu2023-kf, Palos2024-xn, Rashmi2025-vn} for testing purposes. \mbpol\ as a target method is particularly suitable, as it has demonstrated subchemical accuracy compared to CCSD(T) for water clusters while predicting experimental properties (structural, thermodynamic, and dynamical) of water with high fidelity.\cite{Zhu2023-kf, Palos2024-xn} We evaluate this methodology on liquid water, incorporating nuclear quantum effects through converged path integral molecular dynamics (PIMD) simulations.\cite{Marx1996-fp}

\section{Methodology}
\mbpol\ and \mbscan\ (for details, see SI) PBC simulations with a unit cell of 256 water molecules have been carried out using the LAMMPS software\cite{Thompson2022-ep} built with the MBX plugin.\cite{Gupta2025-oq} PIMD simulations were performed using 32 Trotter replicas, with a timestep of 0.25 fs and a local PILE thermostat ($\tau$=\mbox{100 fs}, $T$ = \mbox{298.15 K}).\cite{Ceriotti2010-zb}

MLPs were constructed using the Atomic Cluster Expansion\cite{Drautz2019-sg} with Finnis-Sinclair embedding as implemented in the PACEmaker software.\cite{Lysogorskiy2021-dx} For the \bmlp\ model, the final potential contained 800 functions/element up to body order 6, with a cutoff radius of \mbox{6.0 \r{A}}, and training was done on 100 PBC structures including forces.

\nwater{n} clusters were extracted from PBC simulations by selecting a random water molecule and the $n-1$ nearest water molecules based on oxygen-oxygen distance. \mbpol\ and \mbscan\ settings for cluster single points were in accordance with those in PBC calculations (for details, see SI).

For the \dmlp\ model, the final potential contained 320 functions/element up to body order 3, with a cutoff radius of \mbox{4.0 \r{A}}, and training was done on 1000 \nwater{15} clusters using only energies.

\section{Results and Discussion}

The accuracy of cluster-based learning largely depends on how well the extracted clusters represent bulk data. For sufficiently large clusters, the forces at the cluster center should converge to the corresponding force in the bulk. To quantify this finite size error on the central water molecule, we compared forces calculated on clusters with the corresponding value in the bulk for CC (represented by \mbpol) and SCAN (represented by \mbpol\ and \mbscan).

\begin{figure}[H]
    \centering
    \includegraphics[width=0.95\linewidth]{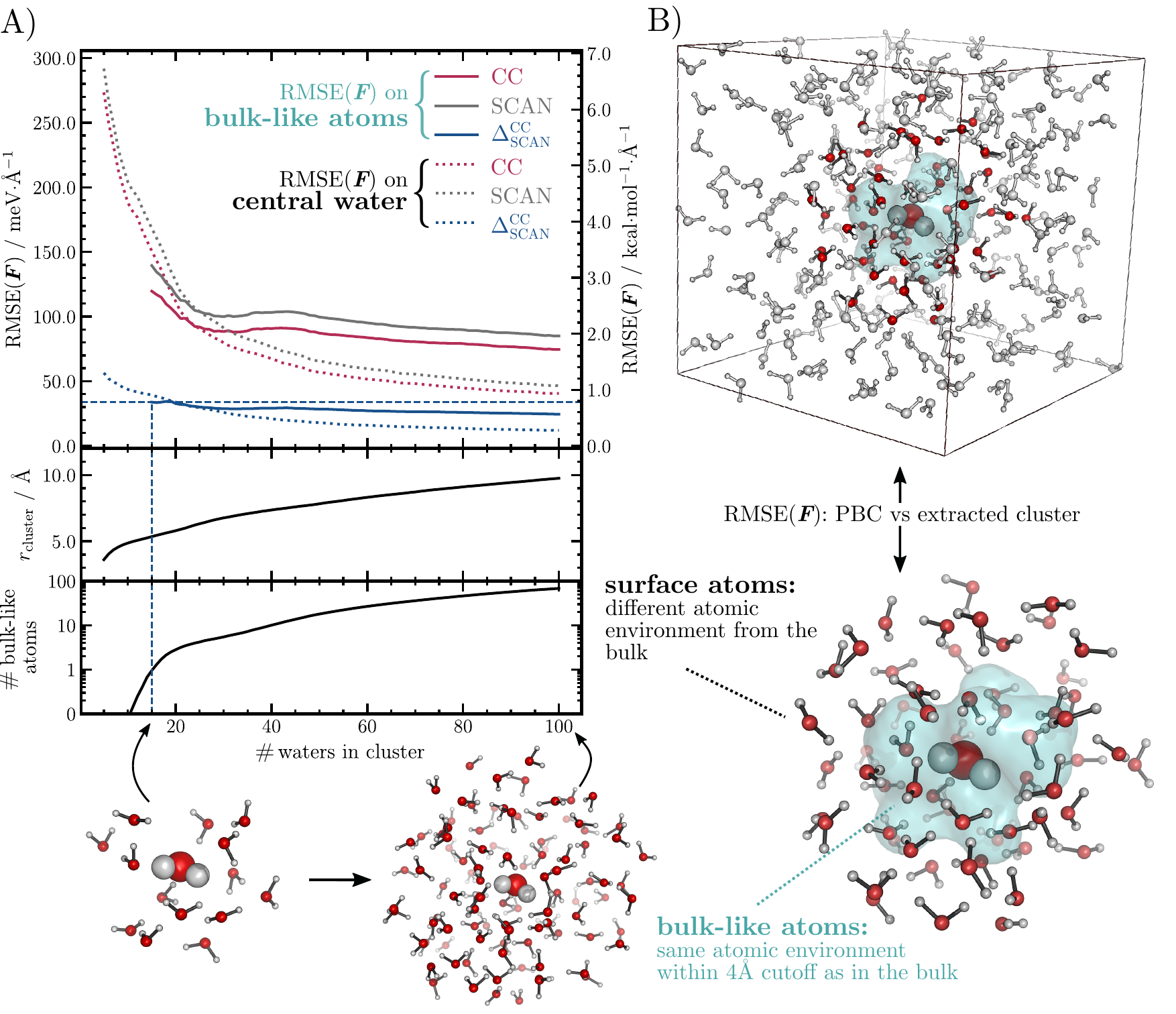}
    \caption{Force analysis in water clusters as a function of size \textbf{A) Top}: RMSE of forces for water clusters (from \nwater{5}\ to \nwater{100}) wrt reference forces calculated for bulk structures under PBC. RMSE($F$) for bulk-like atoms (solid line), and the central water molecule (dotted line). Bulk-like data plotted where at least one bulk-like atom on average (15 water molecules). RMSE($F$) is based on 500 clusters randomly sampled from a 300 ps MB-SCAN PIMD simulation for each cluster size. The horizontal dashed line (33.8 meV/\r{A}) marks the RMSE($F$) for bulk-like atoms in \nwater{15} clusters. \textbf{A) Middle}: Average cluster radius, measured as the distance between the central atom and the furthest atom in the cluster. \textbf{A) Bottom}: Average number of bulk-like atoms in clusters. \textbf{B)} 3D structure of PBC cell and extracted cluster: hydrogens - gray, oxygens in cluster - red, oxygens outside cluster - gray, bulk-like atoms - blue highlight, central water - van der Waals representation}
    \label{fig:cluster-tests}
\end{figure}

Increasing the cluster size by adding more water molecules to the surface leads to a decreasing rate of change in the \rmsef\ error on the central water (Figure \ref{fig:cluster-tests} A). Large initial errors decrease below 100 \mevpera\ by a cluster size of 25. Beyond this point, the decay slows significantly, with errors remaining substantial (50 \mevpera) even for clusters containing 100 water molecules. At this size, the cluster radius approaches \mbox{10 \r{A}} (Figure \ref{fig:cluster-tests} A), indicating that the main source of error is long-range electrostatic interactions. In contrast, the difference between CC and SCAN (\deltac) is dominated by short-range exchange and correlation effects.  Although its decay rate is slower than the original methods, it starts from a significantly lower \rmsef, around 50 \mevpera. Remarkably, at only 15 water molecules, it outperforms the original approach at 100 water molecules, enabling the use of much smaller clusters for training.

To determine the optimal cluster size range for training \dmlp, we also analyzed \rmsef\ on "bulk-like" atoms (Figure \ref{fig:cluster-tests} B)—those with the same atomic environment in the cluster as in the bulk within the MLP cutoff (4.0 \r{A}). This metric is more representative, as the final MLP potential learns from the environments of all bulk-like atoms within its cutoff, not just the central one. Unlike the \rmsef\ on the central water, it does not always decrease monotonically and decays much slower. As a result, for CC and SCAN, \rmsef\ on bulk-like atoms remains significant even at \nwater{100} clusters (over 65 \mevpera). In contrast, for \deltac, it starts at 33.8 \mevpera\ with 15 water molecules. Notably, this error falls below the typical errors of state-of-the-art machine learning models (35-120 \mevpera\ for quantum liquid water\cite{Cheng2024-vb}). Therefore, training the short-range $\Delta$ forces on small clusters offers significant technical and computational flexibility at the cost of negligible and controllable additional error. For larger clusters, the convergence rate is slower for \deltac\, reaching 24.5 \mevpera at 100 water molecules. Therefore, once clusters contain at least one bulk-like molecule, further increasing the cluster size provides only marginal improvement in representing true bulk-like forces.

\end{multicols}

\begin{multicols}{2}

Given that modern MLP frameworks are expected to achieve \rmsef\ on the order of 100 meV/Å,\cite{Omranpour2025-da} direct cluster-based learning requires large clusters to provide accurate enough data. In contrast, the \clusterdelta\ approach enables more efficient training with smaller clusters, which is crucial given the steep scaling of the target high-level methods without losing accuracy. While the original CCMD potential relied on local correlation methods, canonical or non-local approximate CCSD(T) variants also support calculating \nwater{15}\ clusters.\cite{Manna2017-pj, Gyevi-Nagy2021-rc, Sylvetsky2020-yz}

Another advantage of the delta-learning approach emerged when we found that a smaller cutoff and fewer parameters (\mbox{4.0 \r{A}}, 320 functions/element body order 3) were optimal (for details, see SI) compared to the baseline potential (\mbox{6.0 \r{A}}, 800 functions/element, body order 6). Therefore, few-body terms sufficiently capture the \deltac\ difference in water molecule interactions. Given that three-atom terms in ACE can maximally account for three-molecule terms in the many-body (molecule) expansion, this is consistent with the findings of Dasgupta et al. \cite{Dasgupta2021-uh}

\begin{figure}[H]
    \centering
    \includegraphics[width=1\linewidth]{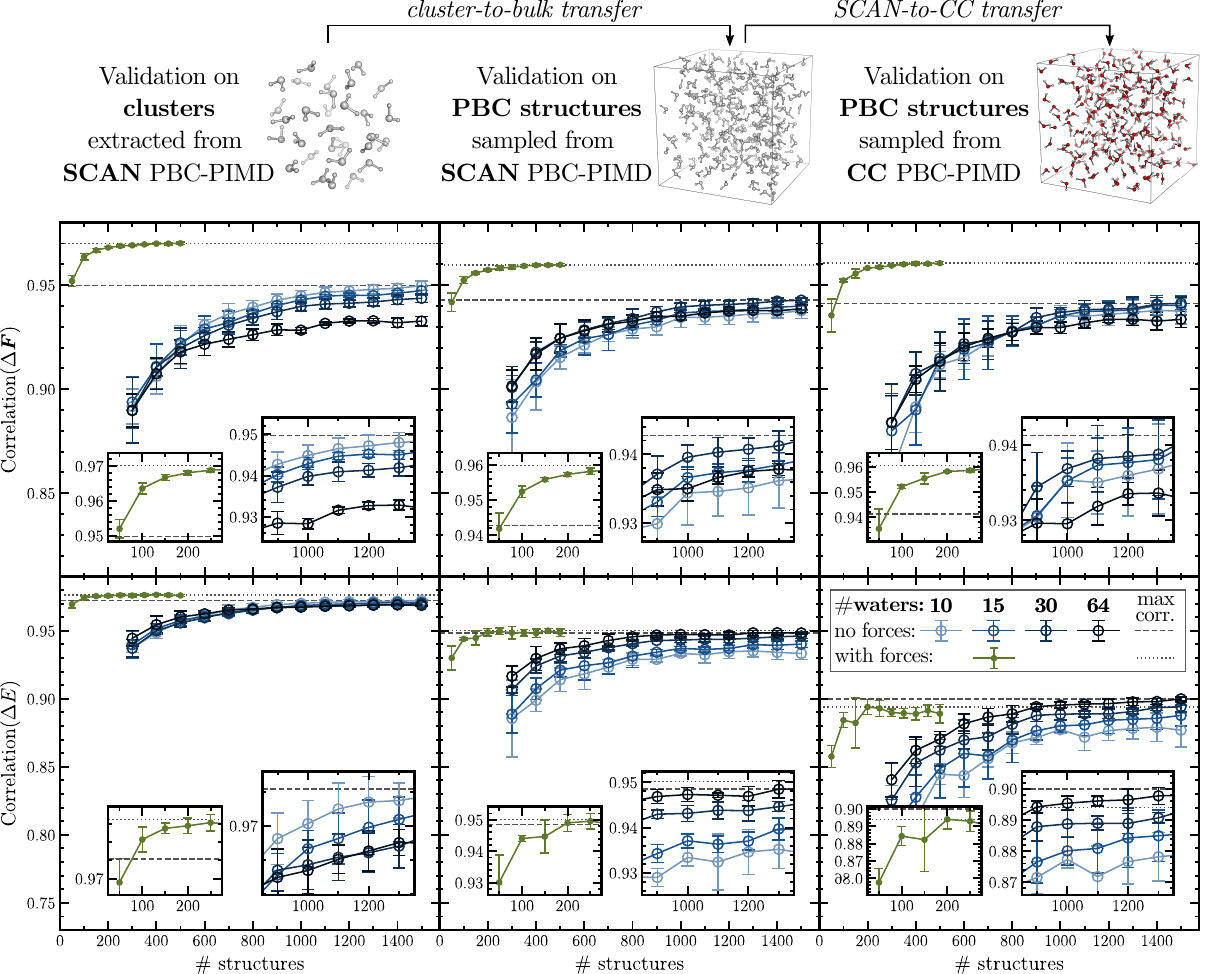}
    \caption{Correlation of \dmlp\ predicted forces and energies as a function of training set size, averaged over five independent fits (using different random seeds, error bars indicate standard deviation). Labels: CC: \mbpol, SCAN: \mbscan. Training data comprised randomly selected clusters from the 0–300 ps interval of SCAN PIMD simulation. Models were trained using only cluster energies (empty circles) or including forces (filled dots, green). Maximum correlations are marked with a dashed line for energy-only training and a dotted line for energy-plus-force training. Validation was performed on: \textbf{Left:} 500 SCAN clusters \textbf{Middle:} 500 SCAN PBC structures \textbf{Right:} 500 CC PBC structures. All structures were randomly extracted from the corresponding simulation's 300-500 ps interval.}
    \label{fig:learning-curves}
\end{figure}

Next, to evaluate the transferability from cluster to bulk, we validated periodic snapshots sampled from SCAN PBC-PIMD (Figure \ref{fig:learning-curves}, middle). Finally, considering the distinct phase spaces spanned by the baseline and target models, we assessed transferability from SCAN to CC using periodic snapshots sampled from CC PBC-PIMD (Figure \ref{fig:learning-curves}, right).

\end{multicols}
\begin{multicols*}{2}

When training with energies alone, learning curves converge above 1000 structures. Surprisingly, different cluster sizes yield similar results, and using large clusters (\nwater{64}) even reduces force correlation. This aligns with the trends in Figure \ref{fig:cluster-tests} A, where increasing the number of waters provides marginal improvements in \deltac\ RMSE($\boldsymbol{F}$). Furthermore, larger clusters do not add more training information, only more input parameters (coordinates) corresponding to one energy value. However, energy correlation shows a small but systematic improvement with larger clusters.

\dmlp\ force predictions exhibit excellent cluster-to-bulk and SCAN-to-CC transferability, proven by the small decrease in force correlation values. While this transferability is weaker for energy correlations, the cluster-to-bulk transfer still suffers less degradation than the SCAN-to-CC transfer. Nonetheless, robust force correlations will enable accurate molecular dynamics simulations.

Although analytical forces are not yet available for local correlation methods, we also explored a hypothetical scenario by including forces in the training set (Figure \ref{fig:learning-curves}, green). This drastically reduces the required training set size by an order of magnitude, with just 100 clusters yielding excellent results. While the quality of energy prediction remains similar to energy-only training, force predictions improve significantly. Therefore, if forces were available in local CC methods, more accurate potentials could be constructed with only 10\% of the training structures. Additionally, the low number of reference points needed may allow the calculation of numerical gradients using high-level methods.

The error metrics for a selected potential trained on 1000 \nwater{15}\ clusters are summarized in Table \ref{tab:error-metrics}. While the correlation values in Figure \ref{fig:learning-curves} are lower than typically expected, this metric indicates accuracy relative to the \deltac\ forces and energies, which are smaller than the total forces and energies. Accordingly, a high correlation is achieved when combining the \deltaace\ correction with the SCAN baseline and comparing it to total CC values. Overall, \deltaace\ delivers state-of-the-art accuracy in force and energy prediction (\rmsef\ = \mbox{66.3 \mevpera}, RMSE($E$) = \mbox{1.46 \mevperatom}).

\begin{table}[H]
\small
\setlength\extrarowheight{2.5pt}
\setlength{\tabcolsep}{5pt}
    \centering
    \begin{tabular}{ l ccc}
    \toprule
        method & \baselineace & \baselineace+\deltaace \\ \midrule
        Reference & SCAN & CC \\ \midrule
        Corr.($\boldsymbol{F}$)  & 0.9998 & 0.9995 \\
        Corr.($E$) & 0.9972 &  0.9977 \\
        \rmsef & 47.0 & 71.0 \\
        RMSE($E$) & 0.97 & 1.23\\
    \bottomrule
    \end{tabular}
    \caption{Error metrics for \deltaace\ (trained on 1000 \nwater{15} clusters) and \baselineace\ calculated as an average of five independent fits for \deltaace. \rmsef\ in \mevpera, RMSE($E$) in \mevperatom. Validated on 500 PBC structures randomly extracted from the 300-500 ps interval of CC PBC-PIMD simulation.}
    \label{tab:error-metrics}
\end{table}

\begin{figure}[H]
    \centering
    \includegraphics[width=1\linewidth]{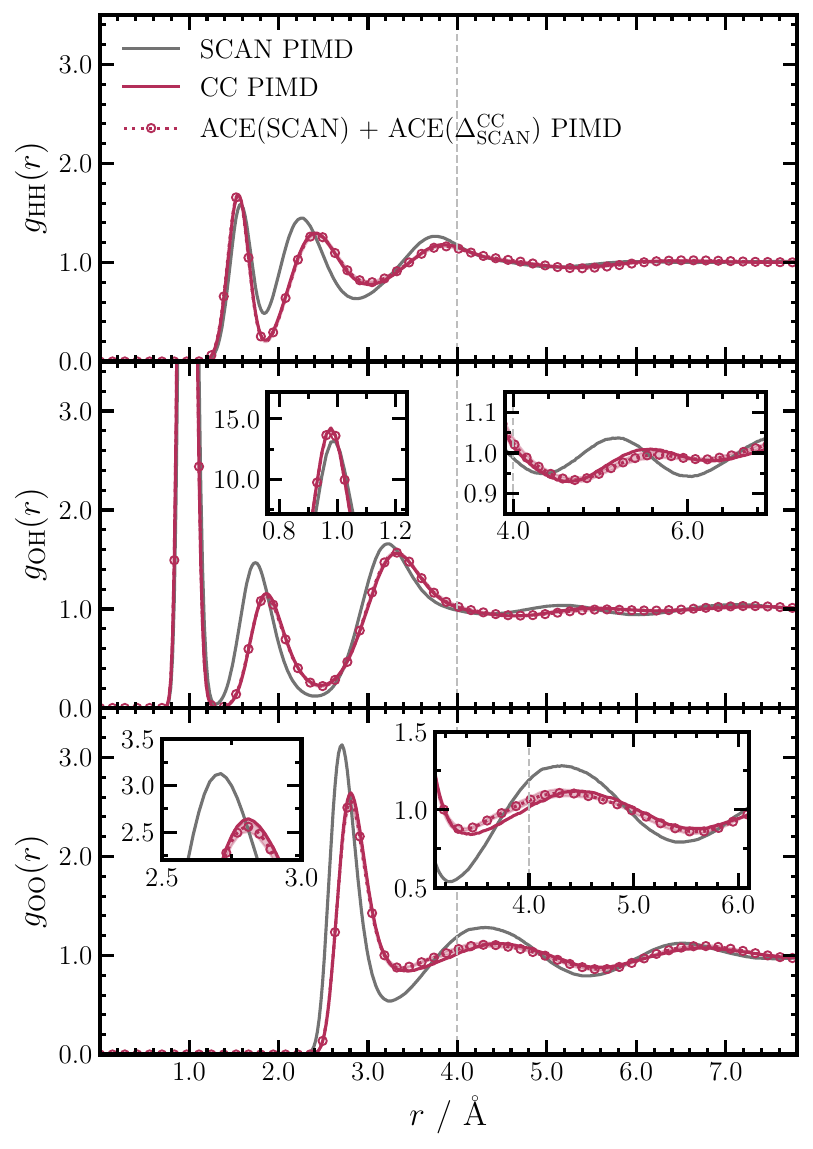}
    \caption{Radial distribution functions of liquid water using different methods, calculated from 200 ps PIMD simulation. Labels: SCAN (\mbox{MB-SCAN}), CC (\mbox{MB-pol}), \baselineace\ (trained on 100 periodic SCAN snapshots), and \deltaace\ (trained on 1000 \nwater{15}\ clusters using exclusively energies). Dotted lines with markers represent the average of five independent fits for the \deltaace\, with shaded areas indicating standard deviation. The vertical dashed line at 4.0 \r{A} marks the cutoff radius for the \deltaace\ potential. RDFs were computed with a 0.01 \r{A} radial bin size and are unsmoothed.}
    \label{fig:gr-deltac}
\end{figure}

Finally, we demonstrated the robustness and accuracy of the \deltaace\ potential in PBC-PIMD simulations. We achieved stable simulations using the \deltaace\ alongside the baseline model (\baselineace), highlighting its strong cluster-to-bulk and SCAN-to-CC transferability. The latter, in particular, represents a significant shift, as evident from the radial distribution functions (RDF, Figure \ref{fig:gr-deltac}); still, accurate simulations are achieved without iterative learning, which was necessary for the original CCMD protocol.

Our \clusterdelta\ strategy yields excellent agreement with the CC reference. The final potential accurately captures the structure of the first two hydration spheres and maintains precise long-range ordering. Remarkably, this is achieved with a \deltac\ MLP cutoff of 4.0 \r{A} in a 19.7 \r{A} PBC cell. Despite each water molecule in the \deltac\ MLP interacting with less than half of the simulation box, it provides an essential contribution to describing the entire periodic system correctly. This reinforces the idea that the dominant force and energy contributions in \deltaace\ stem from short-range interactions, while the \baselineace\ baseline model effectively captures long-range effects.

\begin{figure}[H]
    \centering
    \includegraphics[width=1\linewidth]{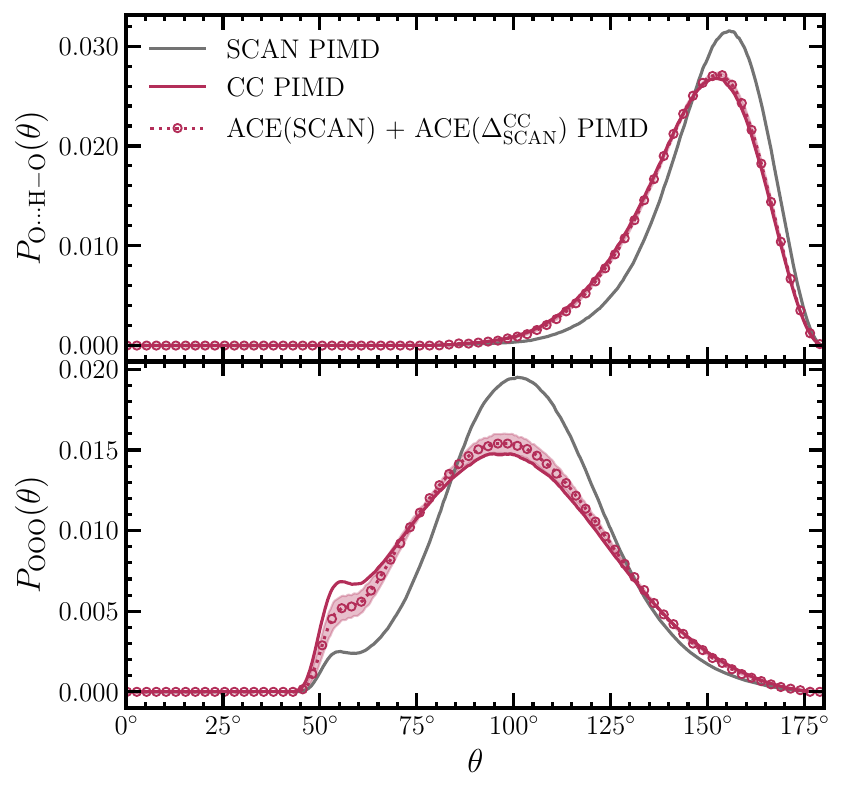}
    \caption{Angular distribution functions (ADFs) of liquid water using different methods, calculated from 200 ps PIMD simulation. $\theta$ is represented in degrees. Labels: SCAN (\mbox{MB-SCAN}), CC (\mbox{MB-pol}), \baselineace\ (trained on 100 periodic SCAN snapshots), and \deltaace\ (trained on 1000 \nwater{15}\ clusters molecules using only energies). Dotted lines with markers represent the average of five independent fits for the \deltaace\, with shaded areas indicating standard deviation. ADFs were computed with an angular bin size of 0.5$^\circ$ and are unsmoothed.}
    \label{fig:adf-deltac}
\end{figure}

Our potential also accurately predicts more sensitive three-body descriptors, such as the angular distribution function (ADF) of hydrogen bonds (O$\cdots$H-O) and oxygen triplets (OOO). As shown in Figure \ref{fig:adf-deltac}, the angle distribution of hydrogen bonds is excellently captured. In the OOO ADF, small deviations toward the baseline potential are observed. The main peak around 100-110° (corresponding to the tetrahedral arrangement) is reproduced well, but the probability of the smaller angled peak is slightly underestimated in the ACE simulations. This structural motif, involving a higher coordination number of interstitial water molecules, is challenging to describe.\cite{Dhabal2014-ea, Sharp2010-hz}

\section{Conclusions}

We propose \clusterdelta, a generally applicable method for generating high-accuracy ML potentials for condensed-phase molecular systems by fitting the discrepancy between high- and low-level force fields using exclusively clusters for the training. This approach retains the simplicity of direct PBC learning while allowing a flexible selection of high-level methods for cluster learning. Our results demonstrate that a moderate set of small molecular clusters (1,000 \nwater{15}\ structures) is sufficient to achieve state-of-the-art accuracy, effectively reproducing structural two- and three-body descriptors. Furthermore, if forces are available for both high- and low-level methods, the required training set size can be reduced by an order of magnitude.

As discussed, our approach's key assumption is the locality of the $\Delta$-forces, which can be directly assessed using periodic or extended cluster reference calculations with analytical or numerical forces. Our analysis shows that the use of exceptionally small clusters is enabled by the fact that $\Delta$-forces converge below an acceptable threshold significantly faster than total forces with respect to cluster size. Given the steep scaling of high-accuracy methods, the reduced system size for reference calculations leads to substantial computational savings. Combined with the need for fewer reference calculations, we forecast 50–200× computational savings for future CCMD training protocols.

\section{Acknowledgements}
The authors are grateful for fruitful discussions with Dominik Marx, Ralf Drautz, and Francesco Paesani.
Financial support for this study was provided by NKFIH grants FK147031, ÚNKP-23-3, DKOP-23, and the ELTE University Excellence Fund. We acknowledge the Digital Government Development and Project Management Ltd. for awarding us access to the Komondor HPC facility based in Hungary.

\printbibliography
\end{multicols*}

\end{document}